\documentclass[a4paper]{article}

\usepackage{INTERSPEECH2022}
\usepackage{xcolor}
\usepackage{todonotes}
\usepackage{comment}
\usepackage{enumitem}
\usepackage{tabularx}
\usepackage[flushleft]{threeparttable} % http://ctan.org/pkg/threeparttable

\title{Data-augmented cross-lingual synthesis in a teacher-student framework}
\name{Marcel de Korte, Jaebok Kim, Aki Kunikoshi, Adaeze Adigwe, Esther Klabbers}
\address{
  ReadSpeaker}
\email{\{marcel.korte,jaebok.kim,aki.kunikoshi,adaeze.adigwe,esther.judd\}@readspeaker.com}

\begin{document}

\maketitle

\begin{abstract}
%\MK{Status: updated (10/03)}
%\MK{Status: incorporated comments (16/03)}
Cross-lingual synthesis can be defined as the task of letting a speaker generate fluent synthetic speech in another language. This is a challenging task, and resulting speech can suffer from reduced naturalness, accented speech, and/or loss of essential voice characteristics. Previous research shows that many models appear to have insufficient generalization capabilities to perform well on every of these cross-lingual aspects. To overcome these generalization problems, we propose to apply the teacher-student paradigm to cross-lingual synthesis. While a teacher model is commonly used to produce teacher forced data, we propose to also use it to produce augmented data of unseen speaker-language pairs, where the aim is to retain essential speaker characteristics. Both sets of data are then used for student model training, which is trained to retain the naturalness and prosodic variation present in the teacher forced data, while learning the speaker identity from the augmented data. Some modifications to the student model are proposed to make the separation of teacher forced and augmented data more straightforward. Results show that the proposed approach improves the retention of speaker characteristics in the speech, while managing to retain high levels of naturalness and prosodic variation.
\end{abstract}
\vspace{0.1cm}
\noindent\textbf{Index Terms}: cross-lingual synthesis, speech synthesis, data augmentation, multilingual synthesis, teacher-student paradigm

\section{Introduction}
%\MK{status: near completion, needs update on key contributions (07/03)}
%\MK{status: incorporated comments (16/03)}

Recent years have brought about vast improvements in the development of TTS systems that are capable of producing synthetic speech that rivals human naturalness \cite{shen2018natural,lancucki2021fastpitch}. The advent of these models has triggered new research outputs related to cross-lingual synthesis, which is the task of synthesizing speech in a language that a speaker is not proficient in, whilst retaining the speaker's identity. 

%\MK{I included this sentence as to highlight the difficulties of cross-lingual synthesis, but if we have a lack of space, we can leave it out} 
%An ideal cross-lingual system performs well on common text-to-speech metrics such as naturalness, intelligibility and robustness, should retain speaker characteristics akin to multi-speaker synthesis, but should in addition either have no accent or controllable accent, retain the prosody of the target language, and where possible it should be able to learn this from oftentimes small quantities of available data. 
Although considerable advancements have been made in this domain \cite{cao2019end,xue2019building,zhang2019learning}, the presented results often highlight a trade-off between maintaining naturalness, preventing accented speech, and retaining key speaker characteristics. This is possibly caused by the fact that models, particularly non-autoregressive models, have insufficient generalization capabilities to separate speaker and language characteristics from each other. As a result, most cross-lingual models are not fully able to generate unseen speaker-language pairs that satisfy all prerequisites. This problem can be overcome to some extent by training models on multiple speakers per language \cite{zhang2019learning,latorre2021combining}, as the presence of multiple speakers can help to implicitly separate speaker and language information. However, not only is this an infeasible approach for many of the world's languages, it also does not fundamentally provide models with the capability to generalize to unseen speaker-language contexts, meaning that it is often not a fully satisfactory solution.

In this paper, we propose to address this generalization problem by applying the teacher-student framework to the cross-lingual task. Concretely, the teacher model is not just used to generate teacher forced data through knowledge distillation such as in \cite{ren2019fastspeech}, but also as a tool to generate free-running augmented data for unseen speaker-language combinations. This augmented data is generated to have access to cross-lingual speech in which speaker characteristics in the unseen language remain present. Including this data into student model training lowers the model's generalization burden, as the desired speaker-language combinations are now no longer unseen. As the augmented data typically has lower naturalness and less prosodic variation than teacher forced data, we made several modifications to the student model that allow it to distinguish between teacher forced data and augmented data.

We show that the proposed approach enables us to generate speech that retains speaker characteristics learned from the augmented data, while maintaining naturalness and prosodic variation levels of models trained on teacher forced data only, and that this improvement can be achieved with negligible effects on latency.

%Concretely, this paper makes the following contributions:
%\begin{itemize}%[noitemsep]
%\item It proposes to apply the teacher-student paradigm to generate augmented data of unseen speaker-language combinations in addition to its common purpose of knowledge distillation.
%\item It proposes to apply the teacher-student paradigm not just for knowledge distillation but also to generate augmented data of unseen speaker-language combinations.
%\item It proposes \EJ{the use of} a duration filter to discard augmented data that suffers from unstable alignments. 
%\item It investigates how the naturalness and prosodic variation present in teacher-forced data can be maintained in the student model after adding augmented data to the model.
%\item It shows that the proposed approach can better retain core speaker characteristics of the target speaker, while maintaining high levels of naturalness and prosodic variation.
%\item It contributes a thorough evaluation of prosodic aspects of cross-lingual synthesis, specifically focused on naturalness of the intonation.
%\MK{An additional benefit is that the student model is faster than the teacher model}

%\item It introduces and evaluates various improvements that can be made to the student model to better distinguish between teacher forced and augmented data.
%\item It shows that with the combination of the above-mentioned, it is possible to attain cross-lingual speech with both high naturalness as well as retention of speaker identity.
%\end{itemize}
%\end{comment}

\section{Related research}
%\MK{status: close to completion (07/03)}
%\MK{status: incorporated comments (16/03)}
% If necessary, I could take away parts of this first paragraph.
% Advances in text-to-speech modeling have led to the introduction of various neural architectures for cross-lingual synthesis.
%However, as each phoneme encoder needs to be trained from scratch, this approach can be problematic for low-resource scenarios. Instead, \cite{tu2019end} proposed to learn a mapping between high-resource source linguistic symbols and low-resource target linguistic symbols, while \cite{staib2020phonological,maniati2021cross} tried to derive phonological features that are common across languages. 
Various approaches have been explored in order to resolve the challenges observed in cross-lingual synthesis. For example, \cite{nachmani2019unsupervised} proposed to use separate phoneme encoders for each language, while~\cite{zhang2019learning} instead chose to use a language embedding that was then added to a large-scale multilingual model. A study on code-switching \cite{cao2019end} evaluated both approaches, and showed that separate encoders can generate cross-lingual speech with higher naturalness and less accent, provided that sufficient data is available to train each encoder. If not, phonological features common across languages can be derived \cite{staib2020phonological,maniati2021cross}. Most of the time, cross-lingual results do highlight a trade-off between producing natural and unaccented speech, and retention of voice characteristics. This is likely caused by insufficient model capabilities to separate speaker and language characteristics. Several papers \cite{zhang2019learning,latorre2021combining,yang2020towards,nekvinda2020one} tried to circumvent this issue by including multiple speakers per language, but the trade-off remained visible \cite{zhang2019learning}, and (partial) loss of speaker characteristics \cite{yang2020towards} or remainders of source language accent \cite{latorre2021combining} were reported in several cases. A recent work \cite{yang2022cross} tried to address this issue and managed to improve speaker similarity ratings while maintaining naturalness levels, but model training became slower and multiple speakers per language were used for the approach.

A data augmentation approach has recently been applied to various text-to-speech contexts \cite{huybrechts2021low,byambadorj2021text,hsu2019disentangling, cooper2020can}. Most related to our work are \cite{xu2020lrspeech}, where data augmentation is applied in a multilingual low-resource setting for the purpose of knowledge distillation, and \cite{hwang2021tts}, where a trained teacher model was used to generate a large corpus of augmented speech with which a student model was trained.
Our setup proposes to extend the latter idea into the cross-lingual domain, where the teacher model is instead used to generate augmented data for speaker-language pairs that are unseen during training of the model.

\section{System architecture}\label{sec:system}
\begin{figure}
    \centering
    \includegraphics[width=\linewidth]{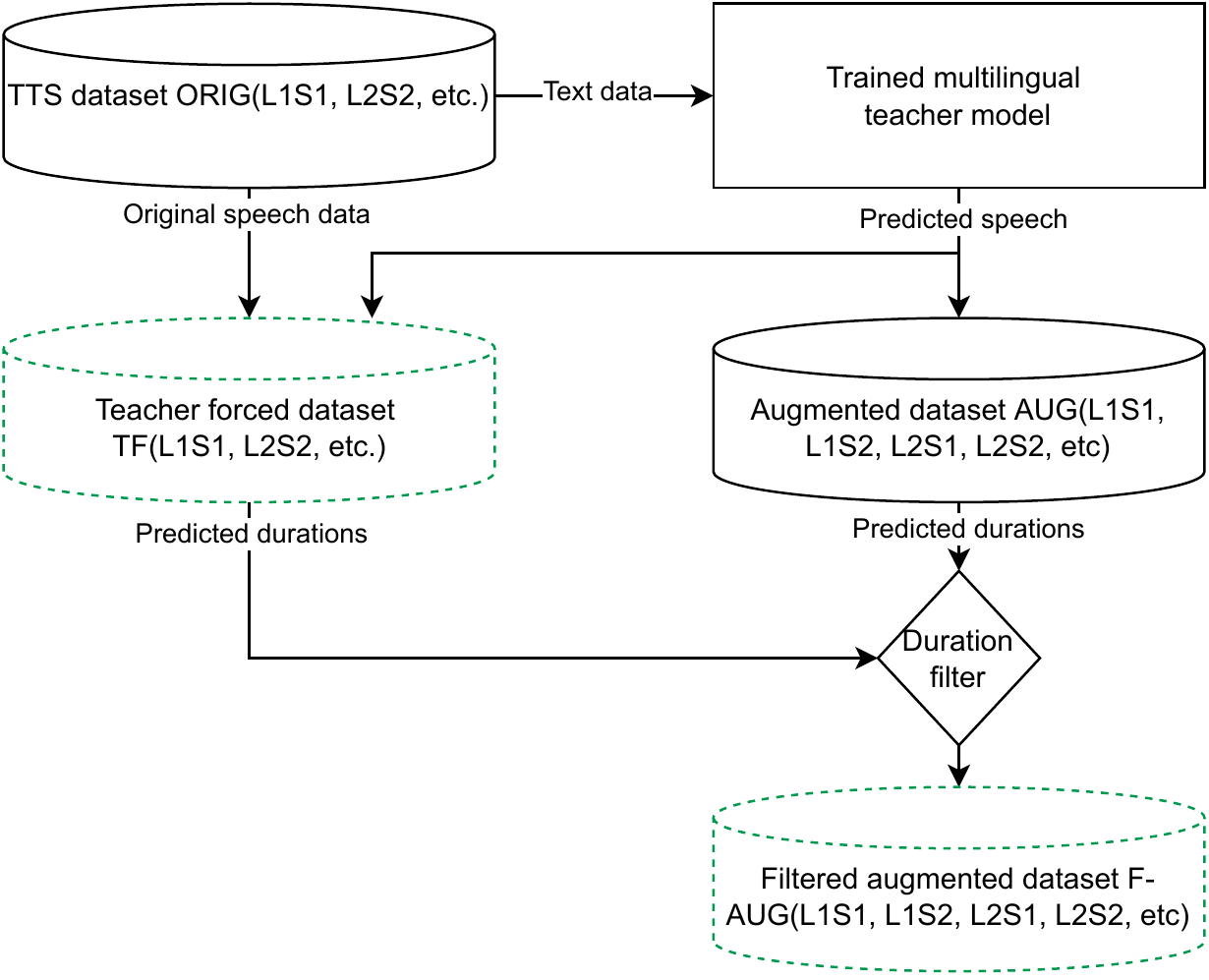}
    \caption{Schematic representation of the proposed data generation approach, where L1 refers to language one, S1 refers to speaker one, etc. Green dotted databases indicate databases that are used for student model training.}
    \label{fig:architecture}
\end{figure}

%\MK{status: close to completion (07/03)}
%\MK{status: comments incorporated (16/03)}
The proposed setup of the data generation process is illustrated in Figure~\ref{fig:architecture}. We first trained a multilingual teacher model and used it to generate 1) a multilingual teacher forced data set obtained through knowledge distillation, and 2) a multilingual augmented data set obtained through data augmentation that contains both in-lingual and cross-lingual data. The data sets are filtered with a duration filter so that sentences with inaccurate alignments are removed. The proposed student model is then trained on both the teacher forced and the augmented data set and a vocoder is used to generate waveforms. The following subsections will describe these steps in more detail.

\subsection{Teacher model}
As a teacher model, we used a modified cross-lingual version of the VoiceLoop~\cite{taigman2017voiceloop} model with dedicated phoneme encoders for each language~\cite{nachmani2019unsupervised}. We refer to~\cite{de2020efficient} for a more detailed description of the model's components.

\subsection{Knowledge distillation and data augmentation}\label{sec:knowledge-distillation}
As displayed in Figure~\ref{fig:architecture}, we used the above-mentioned pre-trained teacher model to generate speech from text. The data set that was used for generating speech was the same that was used to train the teacher model. We refer to this data set as ORIG(L1S1, L2S2, etc.), where L1 refers to language one, S1 to speaker one, etc. We generated two sets of data with the teacher model: 1) multilingual teacher forced data (abbreviated as TF(L1S1, L2S2, etc.)) obtained through knowledge distillation such as in~\cite{ren2019fastspeech}, and 2) free-running augmented data akin to e.g. \cite{guo2019new}. The augmented data is a combination of both seen, in-lingual (AUG(L1S1, L2S2, etc.)), and unseen, cross-lingual (AUG(L1S2, L2S1, etc.)) speaker-language combinations. For each language, the same textual content is used to generate teacher forced, augmented in-lingual and augmented cross-lingual speech.

For the teacher forced data, we used the original output features to condition our current prediction on to ensure equal length and prosodic variation. For the augmented data sets, there is a risk that prediction errors can compound, which can lead to sentences with unstable alignments that can negatively impact the robustness. We therefore introduced a duration filter that first converted the soft alignments into hard alignments, and then removed sentences with unstable alignments or sentences that were both more than 25\% and 30 frames longer or shorter than the original rendering. Applying the duration filter to the data set AUG resulted in a filtered augmented data set F-AUG. This data was then combined with the teacher forced data set TF and used to train the student model with. 

\subsection{Student model}\label{sec:student-model}
An implementation of FastSpeech2~\cite{ren2020fastspeech} was used as the basis for our experiments. For the baseline model, we included a dedicated phoneme encoder per language for cross-lingual modeling, and both an utterance-level encoder for acoustic condition modeling as well as conditional layer normalization to learn speaker information with few extra parameters~\cite{chen2021adaspeech}. This model is able to produce speech with high naturalness but lacks retention of essential speaker characteristics and therefore does not satisfy all demands for cross-lingual synthesis.

As will be shown in Section~\ref{sec:evaluating_pros_var}, the augmented data tends to have less prosodic variation than teacher forced data, which can negatively affect speech generated with the student model. Hence, we made some changes to our proposed model as compared to the baseline model. Concretely, we added an additional embedding that was tasked with learning the difference between teacher forced and augmented data. The embedding output was then included in the duration model and also used to produce an additional scale vector for conditional layer normalization. In addition, batches were reorganized so that sentences from the teacher forced, in-lingual and cross-lingual data with the same phonetic input were part of the same batch.

\subsection{Vocoder}
Waveforms were reconstructed from predicted mel-spectrogram features with a variant of MelGan~\cite{kumar2019melgan}. For the training of the universal vocoder, we used samples of all voices present in the experimental setup. 

\section{Experimental setup}
%\MK{status: near completion (9/3)}
%\MK{status: comments incorporated (17/3)}

Our experiments are divided into two parts. The first part compares the baseline approach trained on teacher forced data with the proposed approach that was trained on teacher forced and augmented data. We investigate whether the proposed model is able to retain speaker characteristics better than the baseline model, while also maintaining high naturalness levels like the baseline approach. The second part analyzes the detrimental effects that augmented data can have on the naturalness and prosodic variation of student model speech, and analyzes how such negative effects can be overcome. Audio samples related to the experiments can be found online\footnote{https://readspeaker-ai-research.github.io/AI-Research/publications/cross-lingual-data-augmentation/}.

\begin{comment}
In this section, we tried to answer the following questions:

\begin{enumerate}
    \item How well does our proposed approach retain core speaker characteristics when compared to the baseline approach trained only on teacher forced data?
    \item How well does our proposed approach using augmented data retain naturalness when compared to the baseline approached only trained on teacher forced data?
    \item How well can our proposed student model retain variation in intonation present in teacher forced data when compared to augmented data? \MK{This question is important to ensure that it does not just learn everything from the augmented data (speaker similarity etc.) but also learns intonation variation present in the teacher forced data}
    \item What effect do the proposed model changes have on the naturalness of the intonation?
    \item What impact does the addition of in-lingual augmented data have on the naturalness of intonation?
    \item How well can our proposed model learn intonation differences between teacher forced and augmented speech?
\end{enumerate}
\end{comment}

\subsection{Data}
We used data from four different speakers, each from a different language. First, a British English female voice was included so that English could function as target language. The remaining three voices were each chosen to have one distinctive characteristic from this voice. We included: a) a female Polish voice, where the language family was different (Slavic vs. Germanic language), b) a male German voice, which differs in sex (male vs. female), and c) a female Swedish voice, where the language accentuation was different (pitch-accent vs. stress-accent). All data was recorded by professional speakers in a studio environment, and then down-sampled to 22 kHz.

\subsection{Training procedure}
For the teacher model training, we randomly selected 10.000 sentences for training and 1.000 sentences for validation for each voice. The model was first pre-trained for 10 epochs and then finetuned for 100 epochs, and trained as described in \cite{de2020efficient}. With this model, we generated teacher forced data for all languages, in-lingual augmented data of all voices in their source language, and cross-lingual data of the Polish, German and Swedish voice in British English. Duration filtering was applied to the augmented speech following the procedure described in Section~\ref{sec:knowledge-distillation}, and about 0.58\% of the data was discarded as a result. We used the teacher forced data set to train a baseline student model as described in section~\ref{sec:student-model} for 500 epochs, and trained the proposed models with all the filtered augmented and teacher forced data for 400 epochs.

\begin{table}
\caption{$F_{0}$ standard deviation of voices in Hz for the English and Polish voice for in-lingual generation.}
\centering
\begin{tabular}{lcc}
\toprule
\textbf{Model} & \textbf{English} &\textbf{Polish}  \\ \midrule
\textit{Original data}  & 29.73 & 33.35 \\
\textit{Teacher - Teacher forced generation} & 29.90 & 33.76 \\
\textit{Teacher - Augmented generation} & 21.23 & 26.22 \\
\midrule
\textit{Student - Baseline TF} & 27.88 & 32.15 \\
\textit{Student - Baseline TF + F-AUG} & 25.70 & 28.57 \\
\bottomrule
\end{tabular}
\label{tab:Experiment_F0_stdev_inlingual}
\end{table}

% Duration figures:
% Cross-lingual: 29555 (1.48% discarded)
% In-lingual: 39855 (0.36% discarded)
% Teacher forced: 39957 (0.11% discarded)

% Cross-lingual: Ania (9909), Maja (9941), Max (9705) -> Total: 29555
% In-lingual: Alice (9967), Ania (9992), Maja (9904), Max (9992) -> Total: 39855
% Teacher forced: Alice (9992), Ania (9978), Maja (9990), Max (9997) -> Total: 39957

% Duration removal procedure:
%We then chose the following conditions to remove sentences with potentially bad alignments: sentences that a) are more than 20% longer or shorter and at least 30 frames longer or shorter than the teacher forced sample, or b) contain phonemes of 0 frames in the augmented alignment which are at least 3 frames in the teacher forced alignment, or c) have more than 2 phonemes with a length of zero frames, or d) predicts the last phoneme with a length of 0 frames, which can indicate that prediction was halted prematurely.

\subsection{Evaluation methods}
The evaluation focused on three different aspects of the cross-lingual speech: speaker similarity, naturalness, and prosodic variation through a new metric that we called naturalness of intonation (see Section \ref{sec:evaluating_pros_var}). 
Speaker similarity was evaluated with an ABX test design. Listeners were presented with a vocoded reference sample of a speaker in its native language, a sample from the baseline model, and a sample from the proposed model. They were asked to choose which of the latter two samples sounded more like the speaker in the reference sample, or if they had no preference. Naturalness was evaluated with a MOS design, using a scale from 1 to 5 with whole point intervals, ranging from 'very unnatural' to 'very natural'.

\subsubsection{Evaluating naturalness of intonation}\label{sec:evaluating_pros_var}
Prosodic variation can be a hard metric to define to participants in a listening experiment. Instead, we decided to focus on intonation, which is one of prosody's most salient aspects and more straightforward to evaluate. Intonation can be defined as the manipulation of the fundamental frequency ($F_{0}$) for communicative or linguistic purposes~\cite{taylor2009text}. We evaluated intonation both subjectively and objectively. The naturalness of the intonation was subjectively evaluated with an AB test design, where listeners were asked which of the two presented samples they thought had more natural intonation, or whether they had no preference. For objective evaluation, we used $F_{0}$ standard deviation as a metric and assumed that a larger standard deviation indicates more natural intonation. To ensure that the computed $F_{0}$ standard deviation values were representing intonation accurately, $F_{0}$ outliers and unvoiced regions were discarded. 

We first evaluated the intonation variation metric on in-lingual data. From Table \ref{tab:Experiment_F0_stdev_inlingual}, it can be observed that the augmented data has substantially lower $F_{0}$ variation than both original and teacher forced data. While a baseline student model trained only on teacher forced data (TF) could reasonably maintain $F_{0}$ variation, scores dropped when filtered augmented data (F-AUG) is also added. The second experiment analyzes whether these effects also occur in cross-lingual synthesis, and evaluates how these effects can be minimized.

\subsection{Listening test design}
We aimed to make our listening test design compliant with the recommendations made in \cite{wester2015we}. We recruited 40 participants, who were randomly assigned one of six batches. Each batch consisted of 18 MOS naturalness screens, 9 ABX speaker similarity screens and 12 AB naturalness of intonation screens, which were randomized within each experiment. The evaluated sentences were between 3 and 5.5 seconds long so that speech fragments would be long enough to evaluate the question asked, but not too long to introduce listener fatigue. For each experiment, three sentences per condition were used per batch. Sentences were reused in other batches, but then generated in one of the other conditions. This way, the text content across batches was similar, minimizing its effects on the results. As such, 18 different sentences were evaluated per experiment. Each part of the listening test was preceded by a demo screen to familiarize participants with the setup. Furthermore, participants were asked about their age group, language background, gender, familiarity with speech technology, and what audio output device they used to allow analysis of inter-group differences.

\begin{table}
\caption{Speaker similarity preference ratings of cross-lingual speech for the target voices in English, synthesized with the Polish, German and Swedish voice.}
\centering
\begin{tabular}{lccc}
\toprule
\textbf{Preference} & \textbf{Polish} & \textbf{German} & \textbf{Swedish} \\ \midrule
\textit{Prefers baseline} & 25.0\% & 1.7\% & 27.5\% \\
\textit{No preference} & 30.0\% & 0.00\% & 34.2\% \\
\textit{Prefers proposed} & 45.0\% & 98.3\% & 38.3\% \\
\bottomrule
\end{tabular}
\label{tab:Experiment_1_a}
\end{table}

\begin{comment}
\begin{table}
\caption{MOS naturalness ratings of cross-lingual speech (standard deviation between brackets)}
\centering
\begin{tabular}{lcc}
\toprule
\textbf{Voice source language} & \textbf{Baseline} & \textbf{Proposed} \\ \midrule
\textit{Polish} & 3.88 (1.00) & 3.59 (1.14) \\
\textit{German} & 1.74 (1.02) & 2.35 (1.03) \\
\textit{Swedish} & 3.93 (1.02) & 3.77 (1.08) \\
\midrule
Average & 3.18 & 3.24 \\
\bottomrule
\end{tabular}
\label{tab:Experiment_1_b}
\end{table}
\end{comment}

\begin{table}
\caption{MOS naturalness ratings of cross-lingual speech  (standard deviation between brackets) in English, synthesized with the Polish, German and Swedish voice.}
\centering
\begin{tabular}{lccc}
\toprule
\textbf{Preference} & \textbf{Polish} & \textbf{German} & \textbf{Swedish} \\ \midrule
\textit{Baseline} & 3.88 (1.00) & 1.74 (1.02) & 3.93 (1.02) \\
\textit{Proposed} & 3.59 (1.14) & 2.35 (1.03) & 3.77 (1.08) \\
%\midrule
%Average & 3.18 & 3.24 \\
\bottomrule
\end{tabular}
\label{tab:Experiment_1_b}
\end{table}

\section{Experimental results}
%\MK{status: near completion (10/3)}
%\MK{status: comments incorporated (17/3)}
\subsection{Evaluating speaker similarity and naturalness}

We first evaluated whether the proposed model can better retain target speaker characteristics than the baseline model. Results are shown in Table~\ref{tab:Experiment_1_a}. As can be observed, the proposed model was preferred over the baseline model for all voices with regards to speaker similarity. The difference was most marked for the German male voice, where the baseline version did not adequately succeed to retain core speaker characteristics. For the female voices, the difference was not as pronounced, but the proposed model was preferred for these voices as well.

Naturalness results are specified in Table~\ref{tab:Experiment_1_b}. An improvement of 0.61 MOS was found for the German male voice, while for the Swedish and Polish voices naturalness slightly dropped with 0.16 and 0.29 MOS respectively. Across voices, a MOS improvement from 3.18 to 3.24 was found. The higher standard deviation spread of the proposed model for the female voices indicates that its samples have more varied levels of robustness, possibly because of the added augmented data. Indeed, the rendering from the proposed model was rated higher for 39\% of sentences for both female voices. Given that the naturalness for both these voices was also above 3.5 MOS, the naturalness scores that were attained can still be considered quite good.

%\MK{Move to discussion?}
%Some participants noted that it was hard to compare speaker similarity with a reference sample in an oftentimes unknown language. With regards to naturalness, some participants noted that they could hear from the voice color that voices were not native speakers, and expected these voices to have an accent, which could have affected naturalness ratings. % This is interesting, as it could suggest an interaction between the perceived naturalness of a voice and the lack of accent a voice has, rather than that a voice is too accented.

\begin{table}
\caption{Preference ratings of naturalness of intonation. The header names refer to the subsections where the experiment is described.}
\centering
\begin{tabular}{lcccc}
\toprule
\textbf{Preference} & \textbf{5.2.1} & \textbf{5.2.2} & \textbf{5.2.3} & \textbf{5.2.4} \\ 
\midrule
Proposed approach & 65.0\% & 48.3\% & 48.3\% & 50.8\% \\
No preference & 14.2\% & 25.0\% & 31.7\% & 31.7\% \\
Alternative approach & 20.8\% & 26.7\% & 20.0\% & 17.5\% \\
\bottomrule
\end{tabular}
\label{tab:Experiment_2}
\end{table}

\subsection{Evaluating naturalness of intonation}
For the second experiment, we evaluated the naturalness of intonation in various contexts. We did not consider speaker similarity in these experiments, as speaker information is treated the same in all models. The Polish voice was chosen to conduct evaluations with, as the Polish language is the most different from the English target language out of the three voices.

\subsubsection{Comparing speech from proposed model with augmented speech}
We first evaluated how speech from the proposed model compares with augmented speech in terms of intonation to evaluate whether the student model does not just learn intonation from the augmented speech. From Table~\ref{tab:Experiment_2}, it can be observed that the naturalness of intonation from the proposed model was preferred in 65.0\% of the cases over that of the augmented speech from the teacher model, which was preferred in 20.8\% of the cases. The objective evaluation in Table~\ref{tab:Experiment_F0_stdev} shows that the $F_{0}$ standard deviation of the proposed student model was 30.08 Hz, while the $F_{0}$ variation of the augmented teacher data was only 24.04 Hz. This decrease in intonation variation is comparable to that of the in-lingual speech in section~\ref{sec:evaluating_pros_var}. We conclude that speech from the proposed model is able to retain the level of intonation variation present in teacher forced data rather than the lower intonation variation from the augmented speech.

\begin{comment}
\begin{table}[!ht]
\caption{Preference ratings of naturalness of intonation, comparing augmented speech from teacher model with speech from proposed student model}
\centering
\begin{tabular}{lcc}
\toprule
\textbf{Preference} &  \\ 
\midrule
Prefers teacher model & 20.8\% \\
No preference & 14.2\% \\
Prefers student model & 65.0\% \\
\bottomrule
\end{tabular}
\label{tab:Experiment_2a}
\end{table}
\end{comment}

\subsubsection{Evaluating the effect of the proposed model changes}
We then evaluated how the exclusion of the proposed generation mode embedding into the architecture impacted intonation. In Table~\ref{tab:Experiment_2}, we can see that the version with the generation mode embedding is preferred in 48.3\% of the cases over the version without this embedding. The objective evaluation of $F_{0}$ variation from Table~\ref{tab:Experiment_F0_stdev} shows a decrease to 25.75 Hz without the embedding, compared to 30.08 Hz when the embedding is included. Thus, the embedding plays an important role in retaining the intonation variation present in the teacher forced speech.

\begin{comment}
\begin{table}[!ht]
\caption{Preference ratings of naturalness of intonation, comparing speech from a basic student model with the proposed student model.}
\centering
\begin{tabular}{lc}
\toprule
\textbf{Preference} &  \\ 
\midrule
Prefers teacher model & 26.7\% \\
No preference & 25.0\% \\
Prefers student model & 48.3\% \\
\bottomrule
\end{tabular}
\label{tab:Experiment_2b}
\end{table}
\end{comment}

\subsubsection{Evaluating the effect of in-lingual augmented data}
We also evaluated whether including in-lingual augmented data makes distinguishing teacher forced and augmented data more straightforward. As Table~\ref{tab:Experiment_2} shows, the model without in-lingual augmented data is preferred in 20.0\% of the cases, while the proposed model with such data is preferred in 48.3\% of the cases. The speech without in-lingual data attains a $F_{0}$ standard deviation of 27.81 Hz, which is lower than the 30.08 Hz from the proposed model. The results thus suggest that the inclusion of in-lingual data can be a helpful means to better distinguish between teacher forced and augmented data. %Another possible explanation is that the model performs better because more data is used for training. However, this seems less likely as the baseline model trained on only teacher forced data does retain high $F_{0}$ variation. \MK{Leave this to the discussion section?}

\begin{comment}
\begin{table}[!ht]
\caption{Preference ratings of naturalness of intonation, comparing speech from proposed model without in-lingual data with proposed model with in-lingual data}\centering
\begin{tabular}{lcc}
\toprule
\textbf{Preference} &  \\ 
\midrule
Prefers teacher model & 20.0\% \\
No preference & 31.7\% \\
Prefers student model & 48.3\% \\
\bottomrule
\end{tabular}
\label{tab:Experiment_2c}
\end{table}
\end{comment}

\begin{table}
\caption{$F_{0}$ standard deviation in Hz of the Polish voice in English for cross-lingual generation.}
\centering
\begin{tabular}{lc}
\toprule
\textbf{Model} & $\textbf{F}_{\textbf{0}}$ \textbf{std} \\ 
\midrule
\textit{5.2.1-5.2.4: Student - Proposed} & 30.08 \\
\textit{5.2.1: Teacher - Augmented data} & 24.04 \\
\textit{5.2.2: Student - Baseline, TF + F-AUG} & 25.75 \\
\textit{5.2.3: Student - No in-lingual augmented data} & 27.81 \\
\textit{5.2.4: Student - Proposed, augmented mode} & 26.00 \\
\bottomrule
\end{tabular}
\label{tab:Experiment_F0_stdev}
\end{table}

\subsubsection{Comparing teacher forced and augmented generation}
Finally, we evaluated whether speech generated with an augmented embedding differs from speech generated with a teacher forced embedding, using the same model. The results from Table~\ref{tab:Experiment_2} show that the speech with the teacher forced embedding is preferred in 50.8\% over the speech with the augmented embedding, while the latter was preferred in 17.5\% of the cases. The objective evaluation from Table~\ref{tab:Experiment_F0_stdev} show a similar pattern, where the $F_{0}$ variation decreases from 30.08 Hz to 26.00 Hz. These results show that the embedding and its integration in the model gives the model the capability to learn the distinction between the two data types and should therefore be included.

\section{Conclusions and Discussion}\label{sec:conclusion}
In this paper, we have proposed to use a teacher model to not just generate teacher forced data, but also augmented data of unseen speaker-language combinations to reduce generalization problems in the student model. Our results show that the proposed approach helps to improve the retention of speaker characteristics, and that the architectural changes that were made in the student model are important to retain naturalness and prosodic variation levels observed in teacher forced data. These results are achieved using just one speaker per language, making the approach feasible for low-resource languages.

We see several opportunities for further research. First, we anticipate that the benefits of distinguishing between teacher forced and augmented data are not limited to the cross-lingual domain. It would also be interesting to experiment with smaller quantities of augmented data to learn speaker characteristics, as training time would decrease. Research by \cite{ribeiro2022cross} suggests that reducing the amount of augmented data may be possible. Finally, we would like to evaluate whether we could increase the naturalness of the cross-lingual speech by using a more powerful teacher model or a voice conversion approach.  
%We foresee several future research directions. First, it could be worthwhile to investigate if less augmented data would be required for training. In this research, we have kept augmented data quantities at a similar level as teacher forced data, but research by \cite{ribeiro2022cross} indicates that smaller quantities of augmented data could be used. Furthermore, we have observed that not just the addition of augmented data but also the capacity of the model to distinguish between data sorts is important, a finding which could be applied to other data augmentation scenarios as well. We additionally note that evaluation metrics specific for both cross-lingual synthesis as well as data augmentation may need to be designed. We have attempted to do this by introducing metrics to evaluate intonational variation, but designing evaluation metrics for accentedness is another aspect that requires attention. Finally, it could be worthwhile to analyze whether naturalness could be further increased by using a more powerful teacher model or by instead opting for a voice conversion approach.

%\section{Acknowledgement}
%We would like to thank xxx
%\MK{Change format}
\pagebreak
\bibliographystyle{IEEEtran}
\bibliography{main}

\end{document}